ESTUDIOS / *RESEARCH STUDIES*

# The influence of online posting dates on the bibliometric indicators of scientific articles

Mercedes Echeverría*, David Stuart**, José-Antonio Cordón-García***

*Universidad Autónoma de Madrid, España
e-mail: mercedes.echeverria@uam.es | ORCID iD: http://orcid.org/0000-0003-4453-5122

**University of Wolverhampton, Reino Unido
e-mail: dp_stuart@hotmail.com | ORCID iD: http://orcid.org/0000-0002-9593-0520

***Universidad de Salamanca, España
e-mail: jcordon@usal.es | ORCID iD: http://orcid.org/0000-0002-8569-9417

**Abstract:** This article analyses the difference in timing between the online availability of articles and their corresponding print publication and how it affects two bibliometric indicators: Journal Impact Factor (JIF) and Immediacy Index. This research examined 18,526 articles, the complete collection of articles and reviews published by a set of 61 journals on Urology and Nephrology in 2013 and 2014. The findings suggest that Advance Online Publication (AOP) accelerates the citation of articles and affects the JIF and Immediacy Index values. Regarding the JIF values, the comparison between journals with or without AOP showed statistically significant differences (P=0.001, Mann-Whitney U test). The Spearman's correlation between the JIF and the median online-to-print publication delay was not statistically significant. As to the Immediacy Index, a significant Spearman's correlation ($r_s$=0.280, P=0.029) was found regarding the median online-to-print publication delays for journals published in 2014, although no statistically significant correlation was found for those published in 2013. Most journals examined (n=52 out of 61) published their articles in AOP. The analysis also showed different publisher practices: eight journals did not include the online posting dates in the full-text and nine journals published articles showing two different online posting dates--the date provided on the journal website and another provided by Elsevier's Science Direct. These practices suggest the need for transparency and standardization of the AOP dates of scientific articles for calculating bibliometric indicators for journals.

**Influencia de la fecha de publicación online de los artículos científicos en los indicadores bibliométricos**

**Resumen:** Este artículo analiza la distancia temporal entre la disponibilidad online de los artículos y su publicación en papel y cómo este lapso de tiempo afecta a los indicadores bibliométricos Factor de Impacto (JIF) e Índice de Inmediatez. Esta investigación examinó 18.526 artículos, la colección completa de artículos y revisiones publicadas por un conjunto de 61 revistas de Urología y Nefrología en 2013 y 2014. Los resultados hallados sugieren que la Publicación Electrónica Anticipada (AOP) acelera la citación de artículos, afectando a los valores de JIF e Índice de Inmediatez. Respecto a los valores de JIF, la comparación entre revistas con AOP y sin esta característica mostró diferencias estadísticamente significativas (P=0,001, U test Mann-Whitney). La correlación de Spearman entre JIF y la mediana del intervalo de tiempo publicación online-papel no resultó ser estadísticamente significativa. En cuanto al Índice de Inmediatez, se halló una correlación de Spearman significativa respecto a la mediana del intervalo de publicación online-papel ($r_s$=0,280, P=0,029) para las revistas publicadas en 2014, pero sin correlación significativa para las publicadas en 2013. La mayoría de las revistas examinadas (n= 52 de 61) publicaron sus artículos en AOP. El análisis mostró también diferentes prácticas de los editores, ocho revistas no indicaron en el texto completo las fechas de publicación online, y nueve revistas publicaron los mismos artículos con dos fechas diferentes de publicación online, la fecha proporcionada por la web de la revista y la fecha suministrada por Science Direct de Elsevier. Estas prácticas sugieren la necesidad de transparencia y estandarización de las fechas de publicación online de artículos científicos para el cálculo de los indicadores bibliométricos de las revistas.

**Palabras clave:** Publicación electrónica anticipada; índice de inmediatez; factor de impacto; fecha de publicación online; tiempos de demora en la publicación; revistas de urología y nefrología.

## 1. INTRODUCTION

The current system for measuring the impact of scholarly journals is based on the quantitative analysis of citations received by scientific publications. Time is an important factor in the calculation of such indicators, and the most popular indicators take into consideration the time a journal has had to accumulate citations, measuring both the citations a journal has received and the time it has taken to accumulate them. However, online publishing is changing the nature of the publication date, with inevitable consequences for the calculating of bibliometric indicators and the ranking of journals.

The most influential indicator to evaluate journals is the Journal Impact Factor (JIF), devised by Garfield and Sher in the early 1960s (Garfield, 2006). A journal's impact factor provides information on the relative importance of a journal in terms of influence, prestige and visibility within its subject category. The JIF is calculated from three elements, citations received, number of substantial items published and the measurement of a two-year window after publication. Despite its apparent accuracy, reproducibility and widespread use, the calculation of these variables as proxy for measuring a journal's influence has been widely discussed (Bordons et al., 2002; Bornmann and Marx, 2016; Glänzel and Moed, 2002; Lozano et al., 2012). The citation window has been criticised for a number of reasons: the shortness of the two-year period for measuring disciplines with different citation latencies, it discriminates against papers whose citation patterns present delayed rises or delayed recognitions (Glänzel and Moed, 2002; Wang, 2013), and the time aggregation for online journal publication escapes from traditional metrics of print journal publication (Haustein et al., 2015; Stuart, 2015; Tort et al., 2012).

The Immediacy Index is another citation metric used for journal evaluation, designed to indicate how quickly articles in a journal are cited. This indicator is calculated as the average number of citations received by articles of a journal in the publication year divided by the number of articles published in that year (Thomson Reuters, 2012). In comparison to JIF, wich is based on citations received by articles published in the two years before the year in question, Immediacy Index is more restrictive in its calculation, considering only citations made in the actual publication year.

This work reports the findings of an investigation into the effect of *online-to-print* delays on the JIF and Immediacy Index of journals categorised as *Urology and Nephrology* in the Journal Citation Reports (JCR). The delay is calculated, from the initial date of online posting provided by the publishers in the full-text of the articles to the final official print date, indexed by Thomson Reuters' Web of Science (WoS). Subsequently, because JIFs form the basis of JCR rankings, we compare the *online-to-print* publication delays across the four quartiles of journals in JCR. Furthermore, the correlation between *online-to-print* delays and the number of articles published by a journal per year is studied, as well as the differences in publication delays between journals published by commercial publishers and scientific societies.

## 2. BACKGROUND

The traditional publication of print journals usually involved long delays from a manuscript's reception to publication. This was due to the bundling of articles into issues, as well as the backlog of manuscripts waiting for placement in a bound print volume with a limited space per year. In the late 1990s the proposal of many publishers to reduce these delays and give visibility to scientific output was to host electronic versions of their paper-based journals on their websites (Dong et al., 2006). Currently, most of major scientific publishers include posting dates in the full-text of peer-reviewed articles, once they are completely edited and formatted, and before the release print issue. For immediate online publication editors use services so-called: *'Advance online publication'* by Nature Group Publishing, *'Advance access'* by Oxford Journals, *'Articles in press'* by Elsevier, *'Early view'* by John Wiley & Sons and *'Online first'* by Springer.

AOP enables the dissemination of knowledge more quickly, reduces the publication print delays and increases the visibility and readership of articles. It might also have influence on bibliometric indicators by enlarging citation time before the print publication date, which is the official start of the citation window used by WoS for calculating the JIF and Immediacy Index.

The increasing *online-to-print* publication delays in recent years, and its influence on bibliometric indicators have been pointed out in a number of recent studies. Tort et al. (2012) showed that the delays between online availability of articles and print publication lead to earlier citations, and thus can artificially raise a journal's impact factor, calculated by WoS on 2-year citation window. Chen et al. (2013) analysed the publication times in 51 *Ophthalmology* journals finding that journals with AOP had a higher JIF compared to those without this feature, although this difference was not statistically significant. Heneberg (2013) reported that online first publication severely affects the calculation of the JIF and Immediacy Index and disadvantages online-only and print-only journals when evaluating them according to these bibliometric indicators.

Others studies have considered the issue of publication delays, Alves-Silva et al. (2016) found that 'online posting', defined as the period spent between the submission of a manuscript and online publication, was significantly and negatively related to JIF in a set of *Ecology* journals. Amat (2008) defined 'publication delay' as the chronological distance between the stated date of reception of a manuscript by a given journal and its appearance in any print issue of that journal, and determined that the effect of online posting on the publisher's site of the whole publication process reduced the publication delay by 29% in a group of *Food Research* journals. In contrast to the assumption that OA journals have shorter publication delays, Dong et al. (2006) showed that the 'publication speed' of journals, defined as the interval of time from the reception of the manuscripts to online publication, was faster for manuscripts published in AOP by Nature Group than those published in OA by Biomed Central. Although OA articles had a greater impact than articles that were not accessible freely online, they estimated that the impact of AOP was still uncertain.

## 3. RESEARCH QUESTIONS

There is a limited amount of research so far into *online-to-print* delays, carried out on different publication times, research practices and different fields, and thus there is a clear need for further studies in more areas. This study contributes to current understanding of the influence of *online-to-print* delays on the JIF and Immediacy Index, through an analysis of a set of 61 journals of *Urology and Nephrology*, indexed in JCR Thomson Reuters. More specifically, the research questions in this paper are:

- What is the correlation between *online-to-print* delays with JIF and Immediacy Index in *Urology and Nephrology* journals?
- Are there any differences between the distribution of journals in quartiles (JCR) and *online-to-print delays?*
- Is there a relationship between *online-to-print* publication delays and the number of articles published per year by a journal?
- Is there any difference in publication delays between journals published by large commercial publishers and scientific societies?

## 4. METHODOLOGY

To gather research data (see online additional material 2013 and 2014 for details) a set of 61 journals out of 77 indexed by JCR (2015), whose subject category was *Urology and Nephrology* were analysed. In order to construct the dataset of journals (see Appendix Table I, and online additional material 2013 and 2014), the following methodological criteria were established:

### 4.1. Criteria for inclusion

- All journals indexed by JCR in 2015 under the category *Urology and Nephrology.*
- All original articles and reviews, whose print publication date was included between January 2013 and December 2014.

### 4.2. Criteria for exclusion

- Electronic journals without print publication version.
- Journals whose articles do not provide information about the online posting dates, JIF or Immediacy Index. Furthermore, those journals that could not be accessed full-text and journals with a low number of articles published were also excluded.
- Book reviews, critiques, comments, and letters from or to editors.
- Journal supplements were not included (following Björk and Solomon, 2013; Chen et al., 2013), as well as monographic series.

All papers were examined in electronic form. When an article was posted with different online dates, the date provided on the journal's website and the date provided by *Elsevier's Science Direct*, the date selected was the first date in which the article was available online. For each article the online posting date, provided by publishers in the full-text online, and the official print date provided by WoS, were recorded (see online additional material 2013 and 2014). When a journal did not specify information about the online posting date it was necessary to contact publishers. This only applied to two journals (*Archivos Españoles de Urología* and *Nephrology Nursing Journal*) and they confirmed that electronic and print publication dates were the same (see Appendix, Table I).

### 4.3. Journals and papers examined

A total of 18,526 articles and reviews published in 2013 and 2014 were examined. This number of articles and reviews corresponded faithfully to 61 out of 77 journals used by JCR to calculate the JIF score of 2015 and Immediacy Index of 2013 and 2014 of the subject category *Urology and Nephrology* Journals. The analysis excluded 16 journals, due to criteria detailed in the list of journals (see Appendix, Table II). This fact slightly altered the number of journals included in the list of 2013 (59 journals) (see Appendix, Table I).

Since only a minority of articles provided the day and month of print publication, the journal issue date was converted to days, based on average unit length. In this way, the date of publication was estimated: 1) as the 15th day of each month for monthly journals, and 2) the last day of the first month for bimonthly journals. In cases of irregular journals (i.e. with 8 or 10 issues per year) the date of publication was distributed among the yearly period (following Amat, 2008). When an article was published in print and online during the same period of time, the number of days was not considered (see online additional material 2013 and 2014).

### 4.4. STATISTICAL ANALYSES

Median and Interquartile Ranges (IQRs, 25% - 75%) were calculated with the variable *online-to-print* publication dates, obtained from the time distance between AOP and print publication for all articles and reviews of each journal. All data were tested for normality distribution and because the measures of the variable *online-to-print* publication dates were not normally distributed non-parametric tests were used.

The Spearman correlation test was used to examine the correlation between the median *online-to-print* publication delays and JIF, Immediacy Index and the size of journals, based on the number of articles published per year. The Mann-Whitney U test was used to compare the JIFs of journals with AOP or without AOP, and also to determine whether there were differences between journals published by commercial editors and scientific societies regarding the variable *online-to-print* publication delays. The Kruskal-Wallis test was used for intra-group comparison of distribution of journals in quartiles with the variable *online-to-print* publication.

In our tests the statistical significance was set at 5% and two-sided. The analyses were performed using the SPSS Statistics software (IBM), version 22.

### 5. RESULTS

The analysis revealed that a high proportion of *Urology and Nephrology* journals published their articles ahead of print 85.2% (n=52 out of 61), with a great dispersion in the lapse of time *online-to-print* publication.

The median *online-to-print* publications was 86 days (IQR, 48-171) for journals published in 2013, and 95 days (IQR, 36-171) for journals published in 2014. The chronological distance between *online-to-print* publication dates showed a wide variability. In 2013, the range varied from journals with no-delays 16.9% (n=10) to journals whose median was more than 300 days 5.1% (n=3), with the most frequent interval being 43-84 days 23.7% (n=14) journals (see Fig. 1).

The chronological distance between *online-to-print* publications in 2014 varied from journals with no delays 14.8% (n=9) to 2 (3.2%) journals whose median was more than 300 days, with the most frequent interval being 169-210 days 18% (n=11) (see Fig. 2).

Regarding the study of the JIF (2015), the comparison between the JIF of journals with AOP and journals without AOP showed statistically significant differences (Mann-Whitney $z$=-3.255, P=0.001). The mean of the JIF for journals with AOP was $\bar{x}$=2.953, and for journals without AOP $\bar{x}$=1.003.

With respect to the correlation between JIF and the median days *online-to-print* publication delays, whose values for 2013 and 2014 were grouped, there was no statistically significant correlation (rs=0.158, P=0.233).

In regard to Immediacy Index, the correlation between median days *online-to-print delays* and Immediacy Index was statistically significant (rs=0.280, P=0.029) for journals published in 2014 (n=61) (see Fig. 3). However, for journals published in 2013 (n=56 with Immediacy Index), there was no significant Spearman correlation (rs=0.117, P=0.391) (see Fig. 4).

In answer to the second research question, about whether there were differences of rank position of journals in quartiles (JCR) regarding *online-to-print* delays, the Kruskal-Wallis test did not identify statistically significant differences in 2013 ($X^2$=1.814, DF=3, P=0.612) and 2014 ($X^2$=2.360, DF=3, P=0.501). However, it is important to observe that journals without AOP were classified in the lower quartiles: 5 titles (Q4) and 4 titles (Q3), whereas 16 journals with 267 days of AOP, on average, were classified in the first quartile (Q1).

Regarding the relationship between *online-to-print* publication delays and the size of journals, in terms of the number of articles published per year by a journal, a significant Spearman correlation was observed, both for journals published in 2013 (rs= 0.335, P=0.010) and 2014 (rs=0.375, P=0.003).

**Figure 1.** Frequency distribution of the median of days online-to-print publication delays in 2013

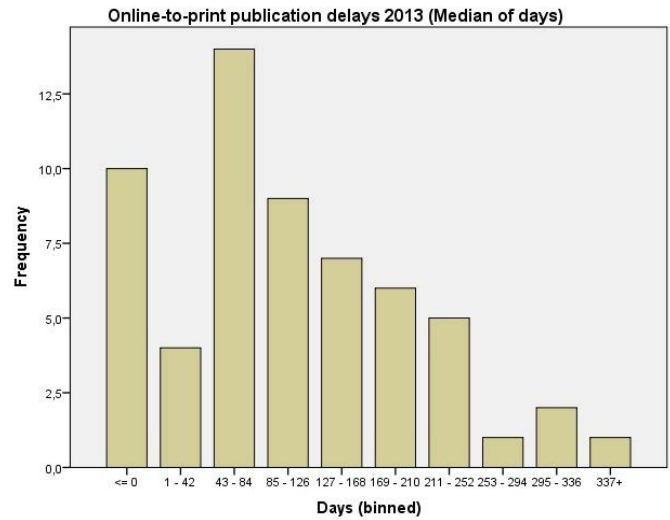

**Figure 2.** Frequency distribution of the median of days online-to-print publication delays in 2014

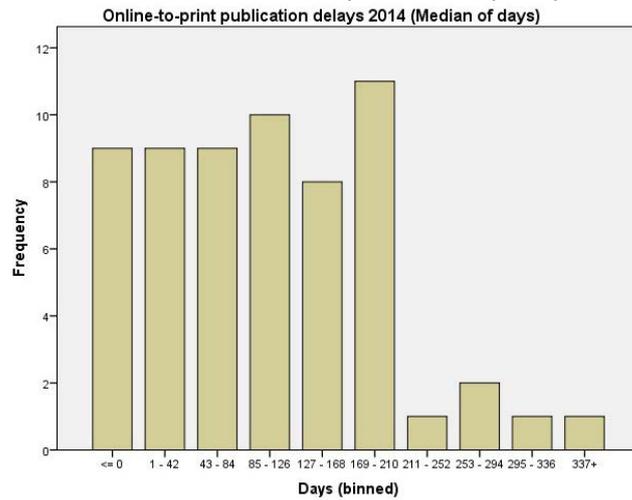

**Figure 3.** Online-to-print publication delays (median of days) and Immediacy Index (2013)

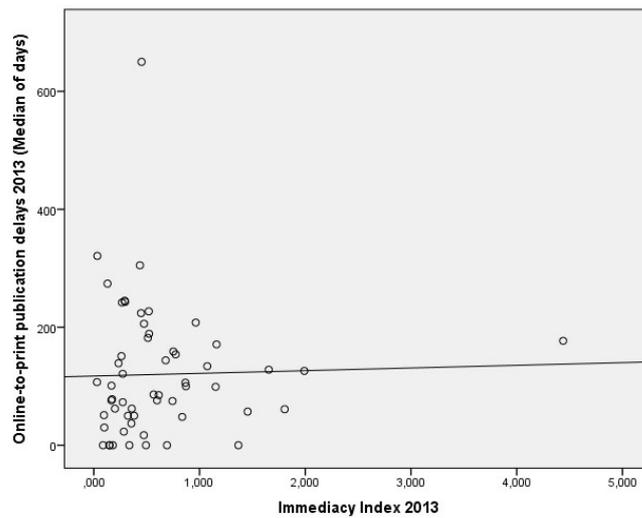

**Figure 4.** Online-to-print publication delays (median of days) and Immediacy Index (2014)

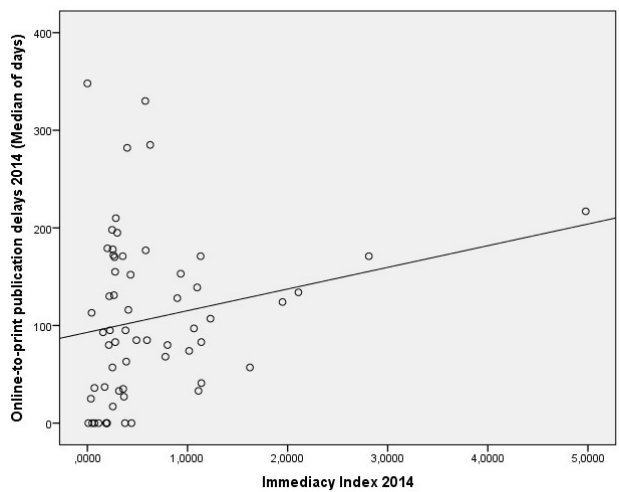

Another issue emerging from this research was whether there were differences regarding publication delays between journals published by commercial publishers (53/61) and by scientific societies (8/61) in relation to the variable of *online-to-print* publication delays. The Mann-Whitney U test identified statistically significant differences for journals published by commercial publishers and scientific societies in 2013 ($z=-2.109$, $P=0.035$, median *online-to-print delays* of publishers 101 days [IQR, 51-173] vs societies 31 days [IQR, 0-81]). For journals published in 2014 ($z=-2.119$, $P=0.034$, median *online-to-print delays* of publishers 97 days [IQR, 45-171] vs societies 34 days [IQR, 0-74]).

## 6. DISCUSSION

The analysis revealed that most publishers of *Urology and Nephrology* journals published articles ahead of print (n= 52 out of 61, 85.2%), with a great dispersion in the lapse time *online-to-print* publications (see Fig. 1 and Fig. 2). The period spent between online to print publication was almost 3 months (86 days), on average, for journals published in 2013 and more than 3 months (95 days), on average, for journals published in 2014.

Although most publishers included the online posting date in the full-text of articles, some exceptions were observed: 8 journals did not include the date of online posting, and 9 journals published the same articles with 2 different online posting dates, the date provided on the journal's website and the date provided by Elsevier' *Science Direct*. These practices reflected a lack of standardization in terms of how publishers report online dates.

Despite the role of time on the accumulation of citations, the analysis did not show significant correlation between JIF and *online-to-print* publication delays. This is probably due to the value of the JIF being determined by different parameters such as the journal size, the subject area, prestige, approach, availability or influential authors, and cannot objectively be reduced to a few variables (Alves-Silva et al., 2016). Althought the extent to which online-to-print delays can affect a journal's JIF is difficult to determine, journals without AOP would have less opportunity to receive citations and consequently it may influence their JIF.

In order to contrast this result to other studies (Chen et al., 2013; Heneberg, 2013; Tort et al. 2012), these authors did not register statistically significant correlation between journals with AOP and JIF, although they mentioned the influence of AOP in terms of higher impact factors in comparison to journals without this feature, number of citations utilized for the calculation of JIF, and potential distortions concerning the evaluation of journals. Moed (2007) provided evidence that the inclusion of a paper in *arXiv* accelerated its citation probably due to this repository making papers available *earlier* accelerating communication, the process of reading and processing of information. This evidence was observed also by Craig et al. (2007) [mentioned by Heneberg (2013)] "preprint availability causes the citation counting process to start earlier although this earlier citation did not affect the final magnitude of citations accrued to a journal article". In this regard, Vanclay (2012) pointed out that "[AOP] offers little advantage for a journal that peaks early (e.g. *Nature*), but may substantially alter the JIF of slow-to-peak journals, such as *Ecology*".

Regarding the Immediacy Index, indicator that provides a useful perspective of the cutting-edge fields of research, the analysis showed a statistically significant correlation with *online-to-print* delays for the journals of 2014 (n=61), although without significant differences for the journals of 2013 (n=57 journals with Immediacy Index). Heneberg (2013), using the database Scopus, studied the number of citations received by articles with AOP ("*in press*" [Elsevier]) within the year when the article was assigned to a journal issue. According to Heneberg (2013), AOP ("*in press*" [Elsevier]) severely affects the calculation of Immediacy Index, although without significant differences.

As Immediacy Index is calculated on the number of citations that an article receives in the same year it is published, an article published early or with AOP has a better chance of being cited than one published later in the year. In fact, as Thomson Reuters (2012) states "Many publications that publish infrequently or late in the year have low Immediacy Indexes". On this point, the potential of technical conditions, such as frequency of publication and publication delays impact the Immediacy Index and have long been recognized and have limited the importance of this Index (Glänzel and Moed, 2002).

In answer to the second research question, about wether there were differences in the distribution of journals in quartiles (JCR) in relation to *online-to-print* delays, the Kruskal-Wallis test did not show any statistically significant differences, although it is observed that a higher median of AOP corresponds with the higher quartiles, and journals without AOP are located in the lower quartiles. However, this statement would need further analysis.

Regarding the correlation between *online-to-print* delays and the number of articles published per year by a journal, the results indicated a statistically significant association. The steady growth of the number of articles, the large backlog of articles awaiting publication in bound print volumes and the need to achieve a rapid and efficient scientific communication are in the origin of Advance Online Publication. In this context, the data reflected statistically significant differences between journals published by major commercial editors (Elsevier, John Wiley, Nature Publishing Group, Oxford, Springer, Taylor & Francis) with a large number of articles per year and long delays versus journals published by scientific societies, with a smaller number of articles published and shorter delays.

It is important to note that commercial editors published 87% of *Urology and Nephrology* journals analysed. As Larivière et al. (2015) pointed out "The proportion of scientific output published in journals under their ownership [large commercial publishing houses] has risen steadily over the past 40 years, and even more so since the advent of the digital era".

## 7. CONCLUSION

The practice of making papers available online ahead of print publication has facilitated the rapid communication of scientific research, but at the same time it has resulted in bibliometric consequences, such as enlarging the citation time of JIF and Immediacy Index.

The research revealed that the longest periods of *online-to-print* publication are mainly produced by large commercial editors that publish journals with a large number of articles per year. This suggests that, to some extent, delays are justified or at least are convenient. However, it would be necessary to study the effects long terms that this gradual increase in the number of days of AOP is having on the number of citations and their contribution to the JIF values, on the differences created between journals with or without AOP, and possible consequences on other bibliometric indicators.

Another feature of some articles with AOP was the lack of standardization of online posting dates. The publishers reported journals with different posting dates for the same article and journals without any mention of the date of online posting. Thus, there is a need for transparency and standardization of posting dates in order to assure the calculation of bibliometric indicators for the evaluation and comparability of journals (Haustein et al., 2015).

Until such standards emerge it is important to have a better understanding of how AOP affects different fields and metrics, and whether there are areas where the *online-to-print* publication delay is sufficient to not only have an impact on lesser-used indicators such as the Immediacy Index but also the widely-adopted JIF.

**APPENDIX**

**Table I.** Dataset of Urology and Nephrology Journals

| Title of journals | 2013 | | | | 2014 | | | | | | Publishers commercial (1) Scientific societies (2) |
|---|---|---|---|---|---|---|---|---|---|---|---|
| | Median days | Interquartile range | Immediacy Index | N articles | Median days | Interquartile range | Immediacy Index | N articles | JIF | Quartiles | |
| Actas Urológicas Españolas | 151 | 95-182 | 0,261 | 92 | 179 | 153-213 | 0,200 | 99 | 0,964 | Q4 | 1 |
| Advance in Chronic Kidney Disease | 151 | 95-182 | 0,261 | 92 | 179 | 153-213 | 0,200 | 99 | 0,964 | Q4 | 1 |
| Aging Male | 50 | 0-60 | 0,323 | 31 | 80 | 57-128 | 0,214 | 42 | 1,493 | Q3 | 1 |
| American Journal of Kidney Diseases | 128 | 100-157 | 1,655 | 219 | 124 | 100-152 | 1,949 | 201 | 6,369 | Q1 | 1 |
| American Journal of Nephrology | 17 | 0-24 | 0,473 | 129 | 33 | 24-46 | 0,315 | 128 | 2,605 | Q2 | 1 |
| American Journal of Physiology-Renal Physiology | 75 | 65-85 | 0,744 | 330 | 68 | 57-77 | 0,779 | 289 | 3,39 | Q1 | 2 |
| Archivos Españoles de Urología | 0 | 0 | No Immediacy Index | 118 | 0 | 0,009 | 0,009 | 113 | 0,307 | Q4 | 1 |
| Asian Journal of Andrology | 76 | 59-102 | 0,598 | 120 | 97 | 61-141 | 1,064 | 125 | 2,644 | Q2 | 1 |
| BJU International | 134 | 64-193 | 1,073 | 424 | 107 | 27-189 | 1,229 | 305 | 4,387 | Q1 | 1 |
| Blood Purification | 0 | 0 | 0,336 | 107 | 0 | 0 | 0,187 | 90 | 1,404 | Q3 | 1 |
| Cardiorenal Medicine | 0 | 0 | 0,179 | 28 | 0 | 0 | 0,194 | 31 | 1,698 | Q3 | 1 |
| Clinical And Experimental Nephrology | 121 | 183-258 | 0,273 | 213 | 282 | 181-301 | 0,397 | 114 | 1,945 | Q2 | 1 |
| Clinical Genitourinary Cancer | 154 | 118-167 | 0,774 | 91 | 172 | 135-199 | 0,26 | 96 | 2,599 | Q4 | 1 |
| Clinical Journal of the American Society of Nephrology | 99 | 77-115 | 1,152 | 252 | 74 | 47-85 | 1,016 | 242 | 4,657 | Q1 | 2 |
| Clinical Nephrology | 139 | 93-414 | 0,234 | 141 | 93 | 47-546 | 0,156 | 102 | 1,065 | Q4 | 1 |
| CUAJ - Canadian Urological Association Journal | 0 | 0 | 1,367 | 231 | 0 | 0 | 0,194 | 218 | 0,866 | Q4 | 2 |
| Current Urology Reports | 63 | 50-82 | No Immediacy Index | 92 | 57 | 34-62 | 0,25 | 84 | 1,597 | Q3 | 1 |
| European Urology | 177 | 143-223 | 4,439 | 189 | 217 | 148-327 | 4,978 | 228 | 14,976 | Q1 | 1 |
| Hemodialysis International | 245 | 204-281 | 0,292 | 87 | 170 | 125-199 | 0,271 | 107 | 1,495 | Q3 | 1 |
| International Journal of Impotence Research | 189 | 140-212 | 0,523 | 43 | 195 | 174-214 | 0,298 | 46 | 1,396 | Q3 | 1 |

| Journal | | | | | | | | | | |
|---|---|---|---|---|---|---|---|---|---|---|
| International Journal of Urology | 208 | 158-236 | 0,964 | 162 | 153 | 130-187 | 0,932 | 221 | 1,878 | Q2 | 1 |
| International Neurourology Journal | 0 | 0 | No Immediacy Index | 32 | 0 | 0 | 0,375 | 32 | 1,344 | Q3 | 2 |
| International Urogynecology Journal | 224 | 187-245 | 0,447 | 286 | 171 | 150-193 | 0,353 | 244 | 1,834 | Q2 | 1 |
| International Urology and Nephrology | 242 | 81-328 | 0,267 | 236 | 155 | 108-184 | 0,277 | 336 | 1,292 | Q3 | 1 |
| Iranian Journal of Kidney Diseases | 0 | 0 | 0,493 | 75 | 0 | 0 | 0,44 | 75 | 1,14 | Q4 | 2 |
| Journal of Endourology | 73 | 41-103 | 0,274 | 237 | 95 | 78-106 | 0,381 | 246 | 2,107 | Q2 | 1 |
| Journal of Nephrology | Articles without online posting in 2013 | | | | 116 | 53-194 | 0,409 | 88 | 1,352 | Q3 | 1 |
| Journal of Pediatric Urology | 274 | 181-391 | 0,130 | 254 | 178 | 139-2015 | 0,253 | 217 | 1,17 | Q4 | 1 |
| Journal of Renal Nutrition | 206 | 76-297 | 0,475 | 80 | 83 | 68-92 | 0,278 | 53 | 2,06 | Q2 | 1 |
| Journal of Sexual Medicine | 85 | 73-125 | 0,614 | 324 | 85 | 75-103 | 0,595 | 315 | 2,844 | Q2 | 1 |
| Journal of the American Society of Nephrology | 61 | 40-87 | 1,805 | 190 | 134 | 114-172 | 2,108 | 260 | 8,491 | Q1 | 2 |
| Journal of Urology | 171 | 146-187 | 1,161 | 606 | 171 | 156-186 | 1,131 | 458 | 4,7 | Q1 | 1 |
| Kidney and Blood Pressure Research | 0 | 0 | 0,087 | 92 | 17 | 0-32 | 0,254 | 71 | 2,908 | Q2 | 1 |
| Kidney International | 126 | 96-153 | 1,991 | 232 | 171 | 143-194 | 2,812 | 242 | 7,683 | Q1 | 1 |
| LUTS - Lower Urinary Tract Symptoms | 321 | 282-361 | 0,033 | 30 | 348 | 298-377 | 0 | 35 | 0,231 | Q4 | 1 |
| Minerva Urologica E Nefrologica | 0 | 0 | 0,148 | 29 | 0 | 0 | 0,048 | 27 | 0,536 | Q4 | 1 |
| Nature Reviews Nephrology | 57 | 47-62 | 1,455 | 55 | 57 | 47-62 | 1,623 | 53 | 9,463 | Q1 | 1 |
| Nature Reviews Urology | 48 | 31-64 | 0,836 | 61 | 33 | 27-42 | 1,109 | 55 | 5,957 | Q1 | 1 |
| Nephrologie & Therapeutique | 107 | 8-153 | 0,03 | 64 | 36 | 0-54 | 0,07 | 51 | 0,541 | Q4 | 1 |
| Nephrology | 23 | 19-26 | 0,284 | 109 | 27 | 24-29 | 0,365 | 126 | 1,796 | Q3 | 1 |
| Nephrology Dialysis Transplantation | 106 | 74-157 | 0,866 | 389 | 139 | 94-187 | 1,096 | 332 | 4,085 | Q1 | 1 |
| Nephrology Nursing Journal | 0 | 0 | 0,148 | 34 | 0 | 0 | 0,111 | 45 | 0,734 | Q4 | 2 |
| Nephron Clinical Practice | 51 | 0-72 | 0,096 | 94 | 37 | 0-53 | 0,172 | 85 | 1,471 | Q3 | 1 |
| Nephron Experimental Nephrology | Only 7 articles in 2013 | | | | 0 | 0-39 | 0,071 | 42 | 1,531 | Q3 | 1 |
| Neurourology Urodynamics | 227 | 207-267 | 0,52 | 97 | 330 | 188-382 | 0,578 | 127 | 3,128 | Q1 | 1 |
| Pediatric Nephrology | 144 | 122-214 | 0,68 | 225 | 177 | 148-270 | 0,582 | 250 | 2,338 | Q2 | 1 |
| Peritoneal Dialysis International | 182 | 150-210 | 0,512 | 85 | 152 | 91-232 | 0,431 | 152 | 1,298 | Q3 | 2 |
| Progrès en Urologie | 101 | 47-137 | 0,167 | 228 | 113 | 71-172 | 0,042 | 212 | 0,562 | Q4 | 1 |
| Prostate | 159 | 108-199 | 0,752 | 182 | 83 | 67-119 | 1,137 | 159 | 3,778 | Q1 | 1 |

| Prostate Cancer And Prostatic Diseases | 100 | 73-131 | 0,871 | 62 | 80 | 53-102 | 0,8 | 55 | 3,803 | Q1 | 1 |
|---|---|---|---|---|---|---|---|---|---|---|---|
| Scandinavian Journal of Urology | 243 | 189-274 | 0,296 | 81 | 198 | 172-222 | 0,247 | 81 | 1,346 | Q3 | 1 |
| Seminars in Dialysis | 86 | 44-190 | 0,566 | 103 | 131 | 77-156 | 0,265 | 113 | 1,912 | Q2 | 1 |
| Seminars in Nephrology | 0 | -37 – 0 | 0,692 | 52 | 41 | 0-47 | 1,138 | 58 | 3,773 | Q1 | 1 |
| Therapeutic Apheresis and Dialysis | 50 | 0-137 | 0,381 | 84 | 210 | 0-248 | 0,284 | 88 | 1,477 | Q3 | 1 |
| Urolithiasis | 62 | 40-95 | 0,200 | 70 | 95 | 69-120 | 0,225 | 71 | 1,454 | Q3 | 1 |
| Urologe | 30 | 0-63 | 0,099 | 163 | 25 | 0-36 | 0,035 | 172 | 0,331 | Q4 | 1 |
| Urologia Internationalis | 76 | 41-115 | 0,166 | 169 | 130 | 81-178 | 0,219 | 160 | 1,313 | Q3 | 1 |
| Urologic Clinics of North America | 78 | 64-95 | 0,173 | 52 | 85 | 66-123 | 0,49 | 51 | 1,835 | Q2 | 1 |
| Urologic Oncology-Seminars and Original Investigations | 650 | 563-744 | 0,452 | 250 | 128 | 90-262 | 0,898 | 283 | 2,921 | Q1 | 1 |
| Urology | 62 | 39-75 | 0,359 | 524 | 63 | 22-78 | 0,388 | 545 | 2,187 | Q2 | 1 |
| World Journal of Urology | 305 | 197-435 | 0,436 | 236 | 285 | 226-312 | 0,627 | 212 | 2,397 | Q2 | 1 |

**Table II.** Journals not included in the analysis

| Journals not included | Reason for non-inclusion |
|---|---|
| Aktuelle Urology | No access available |
| BMC Nephrology | Online only |
| BMC Urology | Online only |
| Canadian Journal of Urology | Articles without date of online posting |
| Contributions to Nephrology | Monographic series (UlrichWeb) |
| Current Opinion in Nephrology and Hypertension. | Articles without date of online posting |
| Current Opinion in Urology. | Articles without date of online posting |
| European Urology Supplements | Supplements |
| International Brazilian Journal of Urology | Articles without date of online posting |
| Kidney International. Supplement. | Supplements |
| Nefrologia | Articles without date of online posting |
| Nephron | JIF not available |
| Nephron Physiology. | Only 11 articles published |
| Renal Failure | Only electronic since 2011 |
| Revista de Nefrologia Dialisis y Trasplante | JIF and Immediacy Index not available |
| Urology Journal | Articles without date of online posting |